\documentclass[final,3p,times,twocolumn]{elsarticle}

\usepackage{amssymb}
\usepackage{graphicx}
\usepackage{amsthm}

\journal{Surface Science}

\begin{document}

\begin{frontmatter}

\title{Strain-induced pseudo-magnetic fields and charging effects on CVD-grown graphene\tnoteref{Acknowledgment}}
\tnotetext[Acknowledgment]{This work was jointly supported by NSF and NRI through the Center of Science and Engineering of Materials (CSEM) at Caltech, and by the Focus Center Research Program-Center on Functional Engineered Nano Architectonics (FENA)}

\author[a]{N.-C. Yeh\corref{cor1}}
\cortext[cor1]{Corresponding author. Tel:+1-626-3954313; Fax:+1-626-683-9060.}
\ead{ncyeh@caltech.edu}
\author[a]{M.-L. Teague}
\author[b]{S. Yeom}
\author[b]{B. L. Standley}
\author[a]{R. T.-P. Wu}
\author[b]{D. A. Boyd}
\author[b]{M. W. Bockrath}

\address[a]{Department of Physics, California Institute of Technology, Pasadena, California 91125, USA}

\address[b]{Department of Applied Physics, California Institute of Technology, Pasadena, California 91125, USA}

\begin{abstract}
Atomically resolved imaging and spectroscopic characteristics of graphene grown by chemical vapor deposition (CVD) on copper are investigated by means of scanning tunneling microscopy and spectroscopy (STM/STS). For CVD-grown graphene remaining on the copper substrate, the monolayer carbon structures exhibit ripples and appear strongly strained, with different regions exhibiting different lattice structures and electronic density of states (DOS). In particular, ridges appear along the boundaries of different lattice structures, which exhibit excess charging effects. Additionally, the large and non-uniform strain induces pseudo-magnetic field up to $\sim 50$ Tesla, as manifested by the DOS peaks at quantized energies that correspond to pseudo-magnetic field-induced integer and fractional Landau levels. In contrast, for graphene transferred from copper to SiO$_2$ substrates after the CVD growth, the average strain on the whole diminishes, so do the corresponding charging effects and pseudo-magnetic fields except for sample areas near topological defects. These findings suggest feasible nano-scale ``strain engineering'' of the electronic states of graphene by proper design of the substrates and growth conditions. 
\end{abstract}


\end{frontmatter}

\section{Introduction}
\label{sec1}

The superior physical properties of graphene~\cite{CastroNeto09} and its compatibility with two-dimensional lithographic processes have stimulated intense research of graphene-based electronics for ``beyond Si-CMOS'' technology~\cite{Novoselov05,Geim07,Miao07,ZhangYB05,Standley08,Tombros07,WangX08}. One of the major challenges for realizing the graphene-based beyond Si-CMOS technology is the fabrication of high-quality large-area graphene sheets and the retention of superior electronic characteristics of graphene. The apparent degradation of carrier mobility and significant variations in the electronic characteristics when graphene comes in contact with various dielectrics~\cite{TeagueML09,RutterGM07} suggest significant susceptibility of the single-layer carbon atoms to the surrounding environment. On the other hand, the strong susceptibility of graphene to external influences also provides opportunities for engineering unique properties of graphene. For instance, it has been theoretically proposed that a designed strain aligned along three main crystallographic directions induces strong gauge fields that effectively act as a uniform magnetic field on the Dirac electrons~\cite{Guinea10a,Guinea10b}. In particular, for a finite doping level, the quantizing field results in an insulating bulk and a pair of counter-circulating edge states, which is similar to the case of topological insulators~\cite{Kane05,Bernevig06,Moore07}. Moreover, strained superlattices have been shown theoretically to open significant energy gaps in the electronic spectrum of graphene~\cite{Guinea10a}, which could be much more effective than current approaches to the bandgap engineering of graphene by either controlling the width of graphene nano-ribbons~\cite{Nakata96,SonYW06} or applying magnetic fields to bi-layer graphene~\cite{McCann06a,McCann06b,Novoselov06}. Recent scanning tunneling microscopic and spectroscopic (STM/STS) studies of graphene nano-bubbles on Pt(111) single crystal substrates~\cite{LevyN10} have verified the theoretical prediction of strain-induced pseudo-magnetic fields, and significant field values in excess of 300 Tesla have been observed.  

In the context of synthesis of large area graphene, various growth techniques have been developed to date, including ultra-high-vacuum annealing to desorption of Si from SiC single crystal surfaces~\cite{Berger06,Emtsev09}, deposition of graphene oxide films from a liquid suspension followed by chemical reduction~\cite{Stankovich08,Eda08}, and chemical vapor deposition (CVD) on transition metals such as Ru~\cite{Sutter08,deParga08}, Ni~\cite{Reina09,KimKS09}, Co~\cite{Ueta04}, Pt~\cite{Starr06,Vaari97} and Cu~\cite{LiX09}. While samples as large as centimeter squared have been achieved, and characterizations of macroscopic physical properties such as Raman scattering and mobility measurements have been reported~\cite{Reina09,KimKS09,LiX09}, most large-area samples appear to exhibit thicknesses ranging from one to approximately twelve layers. Moreover, few investigations with the exception of Refs.~\cite{LevyN10,YehNC10} have been made on correlating the local electronic properties with the microscopic structural variations of these large-area films. 

In this work we report scanning tunneling microscopic and spectroscopic (STM/STS) investigations of the microscopic electronic and structural properties of CVD-grown graphene on copper and also on SiO$_2$ after transferred from copper. We find that the CVD grown graphene is strongly strained so that both the lattice structure and the local electronic density of states (DOS) of graphene are significantly affected~\cite{YehNC10}. The non-trivial strain results in significant pseudo-magnetic fields $B_s$ up to $\sim 50$ Tesla, as manifested by the DOS peaks at energies corresponding to quantized Landau levels $|E_n| \propto \sqrt{n}$, where $n$ denotes both integers and fractional numbers at 0, $\pm 1, \pm 2, \pm 3, 4, 6, \pm 1/3, \pm 2/3, \pm 5/3$. In addition, significant charging effects are found along the strongest strained areas of the sample, consistent with strain-induced scalar potential. On the other hand, for CVD-grown graphene transferred from copper to SiO$_2$, both the pseudo-magnetic fields and charging effects diminish in most parts of the sample except for small regions containing topological ridges that cannot be relaxed by changing the substrate. The occurrence of conductance peaks at quantized integer and fractional Landau levels in strained graphene is analogous to the magnetic field-induced integer and fractional quantum Hall effects (IQHE and FQHE), suggesting that the two-dimensional Dirac electrons may be quantum confined into correlated many-body states by strain. The strained-induced charging effects and quantized states also enable new possibilities of controlling the energy gaps and doping levels of graphene through nano-scale ``strain engineering''. 

\section{Experimental}
\label{sec2}

Our experimental approach to investigating the local electronic and structural correlations of graphene on dielectric or metallic substrates is to perform STM/STS studies using a home built cryogenic STM, which was compatible with high magnetic fields and also capable of variable temperature control from room temperature to 6 K, with a vacuum level of $\sim 10^{-10}$ Torr at the lowest temperatures. For studies reported in this work, the measurement conditions were at 77 K under high vacuum ($< 10^{-7}$ Torr) and in zero magnetic field. The STM tips were made of Pt/Ir alloy and were produced by mechanical cleavage and electrochemical polishing.  The tips thus prepared were first characterized on graphite for atomic sharpness and then cleaned by gently touching the STM tips to an amorphous gold surface. Both topographic and spectroscopic measurements were performed simultaneously at every location in a ($128 \times 128$) pixel grid. At each pixel location, the tunnel junction was independently established so that the junction resistance of 1.5 G$\Omega$ was maintained across the sample. 

The differential conductance, ($dI/dV$), was calculated from the best polynomial fit of each tunneling current ($I$) vs. bias voltage ($V$) curve. The data acquisition typically involved averaging ten $I$-vs.-$V$ spectra per pixel. Polynomial fitting to successive segments of the averaged $I$-vs.-$V$ curve at each pixel was made over a finite voltage range, and the $(dI/dV)$-vs.-$V$ spectrum per pixel was obtained by collecting the derivatives of the best fitted segments at all voltages. The spectra thus derived are generally consistent with those obtained by superposing a modulated bias voltage with a lock-in amplifier on the $I$-vs.-$V$ measurements. However, the modulated-bias technique generally requires much longer time to collect both the spectroscopic and topographical data pixel by pixel over extended areas with high spatial resolution, and so was not employed in this work.  

The procedures for CVD growth of graphene on copper foils were similar to those described in literature~\cite{LiX09}. Briefly, graphene films were primarily grown on 25 $\mu$m thick, highly polished polycrystalline copper foils in a furnace consisting of a fused silica tube heated in a split tube furnace. The fused silica tube loaded with copper foils was first evacuated and then back filled with pure hydrogen gas, heated to $\sim 1000^{\circ}$C, and maintained under partial hydrogen pressure with argon. A flow of methane was subsequently introduced for a desired period of time at a total pressure of 500 mTorr. Finally, the furnace was cooled to room temperature, and the copper foils coated with graphene appeared brighter relative to the as-received copper foils, consistent with previous reports~\cite{LiX09}. The graphene sheets thus prepared were found to be largely single- or double-layered based on Raman spectroscopic studies. Before the STM measurements, a graphene-coated copper foil was first cleaned and then loaded onto the STM probe. The STM probe was then evacuated to high vacuum condition ($< 10^{-7}$ Torr). Two different CVD-grown graphene samples on copper were studied at 77 K. Both topographic and spectroscopic measurements on single-layer graphene regions were carried out on both samples from relatively large area surveys down to atomic-scale investigations. 

To transfer CVD-grown graphene samples to SiO$_2$ substrates, SiO$_2$ substrates were prepared by first thermally grown a 290 nm thick SiO$_2$ layer on the silicon wafer (p-type $\langle 100 \rangle$), followed by gentle sonication of the substrate in acetone and then pure alcohol for about two minutes. The substrate was then baked at 115$^{\circ}$C on a hotplate, nitrogen blown dry while being cooled down. For the CVD-grown graphene on copper, a layer of PMMA was first deposited on top of the graphene sample as scaffolding, and then the copper substrate was removed with nitric acid. Next, the PMMA/graphene sample was placed on a SiO$_2$ substrate with the graphene side down. Finally, the PMMA was removed with acetone. The transferred graphene was cleaned and annealed in an argon atmosphere at 400$^{\circ}$C for 30 minutes. Finally, electrodes were created by thermally evaporating 2.5 nm chromium and 37.5nm gold through an aluminum foil shadow mask. Prior to STM measurements, the sample was rinsed again gently with acetone followed by pure alcohol to remove possible organic surface contaminants. The STM/STS studies on one of the transferred graphene samples on SiO$_2$ were all conducted on the single-layer region.   

\section{Results and Analysis}
\label{sec3}

Systematic studies of the CVD grown graphene on copper revealed large height variations everywhere, as exemplified by the topographic images Figs.~1a - 1c and the corresponding height histograms in Figs.~1d - 1f for one of the two samples. The ripple-like height variations are consistent with previous reports~\cite{LiX09} and are primarily associated with two physical causes. The first is the large difference in the thermal contraction coefficients of graphene and copper. Namely, upon cooling the sample from $\sim 1000^{\circ}$C to room temperature after the CVD growth of graphene, the expanded graphene (due to its negative thermal contraction coefficient) can no longer remain flat on the thermally contracted copper foil, thus resulting in large ripples. The second cause is the original surface roughness of the copper foils, as manifested in Fig.~2, where the surface morphology imaged by an atomic force microscope (AFM) and the corresponding height histograms of a copper foil processed under the same conditions as those for the CVD-grown graphene on copper samples are shown is Figs.~2a - 2d. Comparing Figs.~1 and 2 suggesting that the long-wavelength height variations ($\Delta z$) of the CVD-grown graphene on copper may be associated with the surface morphology of the copper foil, and the averaged variation over a length scale $L$ is estimated at $(\Delta z/L) < 0.01$. On the other hand, significant short-wavelength height variations, such as those exemplified in Figs.~2e - 2f over a small $(3.0 \times 3.0)$ nm$^2$ area may be primarily attributed to the difference in the thermal contraction coefficients, where the height variations could reach as much as 1 nm over a $(3.0 \times 3.0)$ nm$^2$ area for both samples so that $(\Delta z/L) \sim 0.3$, which are much larger than those associated with the long-wavelength height variations and therefore suggest significant strain on graphene.

\begin{figure}
  \centering
  \includegraphics[width=3.2in]{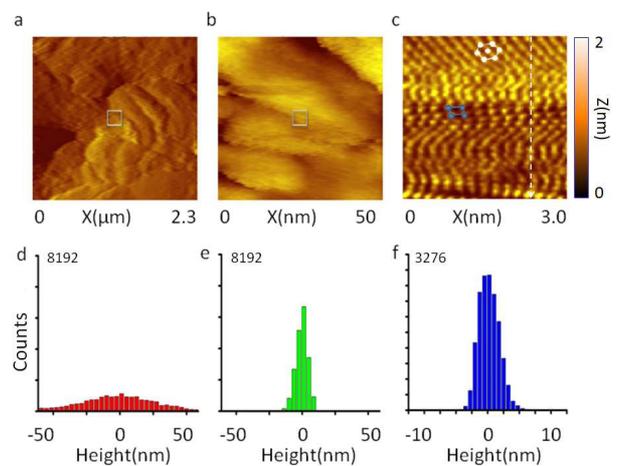}
\caption{(Color online) Topographic images of a CVD-grown graphene-on-copper sample for (a) a $(2300 \times 2300)$ nm$^2$ area, (b) a $(50 \times 50)$ nm$^2$ area and (c) a $(3.0 \times 3.0)$ nm$^2$ area. The white box in (a) represents the region that is enlarged into the image in (b), and the white box in (b) represents the area that is enlarged into the image in (c). The corresponding height histograms for the images shown in (a) - (c) are given in  (d) - (f).}
\end{figure}

The large strain associated with the CVD-grown graphene-on-copper samples apparently incurred significant lattice distortion to the arrangements of carbon atoms. As exemplified in the atomically resolved topographic image in Fig.~3a, different atomic arrangements were apparent in different regions of the $(3.0 \times 3.0)$ nm$^2$ area in view. Specifically, we found that in a more relaxed region of the graphene sample (denoted as the $\alpha$-region) the atomic structure appeared to resemble the honeycomb or hexagon-like lattice with slight distortion. On the other hand, for spectra taken on the areas slightly below the ridge ($\beta$-region) of a ripple-like feature, the atomic arrangements deviated strongly from those of graphene, showing either nearly square-lattice or disordered atomic structure in the $\beta$-region. Additionally, we found certain areas that were completely disordered without atomic resolution, as exemplified in Fig.~3c and denoted as the $\gamma$-region. The disordered region might be associated with amorphous carbon grown on defected features ($e.g.$ the bright features shown in the AFM images in Figs.2a and 2b) on the copper substrate. 

\begin{figure}
  \centering
  \includegraphics[width=3.2in]{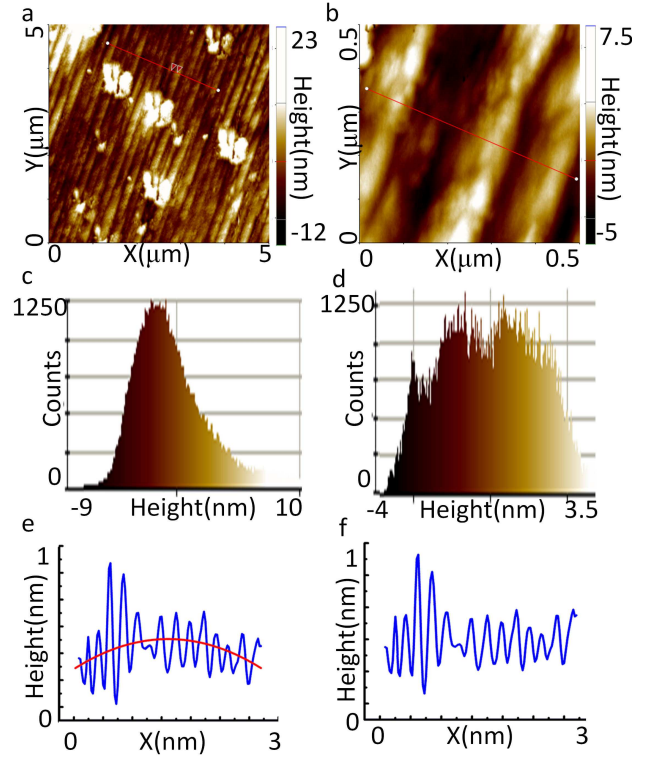}
\caption{(Color online) AFM studies of the surface morphology of a copper foil processed in the same way as the CVD-grown graphene on copper is compared with the STM surface topography of a CVD-grown graphene on copper: (a) An AFM image of the surface morphology of the copper foil over a $(5.0 \times 5.0) \mu {\rm m}^2$ area. (b) An AFM image of the surface morphology of the copper foil over a $(500 \times 500) {\rm nm}^2$ area. (c) The height histogram of the surface morphology along the red line in (a). (d) The height histogram of the surface morphology along the red line in (b). From (a) - (d) we find relatively long-wavelength height variations on the order of $(\Delta z/L) < 0.01$, where $L$ denotes the lateral length for total height variation $\Delta z$. (e) The height variations along the dashed line in the STM topography in Fig.~1c for a CVD-grown graphene on copper, showing atomic steps (solid lines) superposed on top of a gradual background (dotted lines). (f) The height variations in (e) after removing the long-wavelength background, showing sharp steps of either $\sim 0.3$ nm or $\sim 0.6$ nm in height, comparable to one or two c-axis lattice constants of graphite. From (e)-(f) we find short-wavelength height variations on the order of $(\Delta z/L) \sim 0.3$, much larger than the long wavelength variations shown in (a) -- (d).}
\end{figure}

To investigate the effect of lattice distortion on the local electronic DOS of graphene, we compared the typical tunneling conductance ($dI/dV$) vs. $V$ spectra obtained from different regions of atomic arrangements. As shown in Figs.~3b and 3d, the representative spectra for the more relaxed $\alpha$-region of the graphene sample shown in Fig.~1 resembled those found in mechanically exfoliated graphene except a substantially larger zero-bias conductance and additional conductance peaks. On the other hand, the typical spectra associated with the strongly strained $\beta$-region appeared to be asymmetric relative to those in the $\alpha$ region, with sharp conductance peaks more closely spaced. Additionally, for disordered $\gamma$-regions without atomic resolution, nearly parabolic spectra that deviate fundamentally from the linear energy dependence of the Dirac fermions were found, as exemplified in Fig.~3d. For $\alpha$- and $\beta$-type spectra exemplified in Fig.~3b, the energy interval between consecutive conductance peaks appear to decrease with increasing energy. On the other hand, for $\gamma$-type spectra, the conductance peaks seem randomly spaced. Similar results are confirmed in the second CVD-grown graphene-on-copper sample, as exemplified in Figs.~3e - 3f, where the conductance peaks of the representative point spectra correspond to a pseudo-magnetic field $B_s = 35 \pm 5$ Tesla.    

\begin{figure}
  \centering
  \includegraphics[width=3.3in]{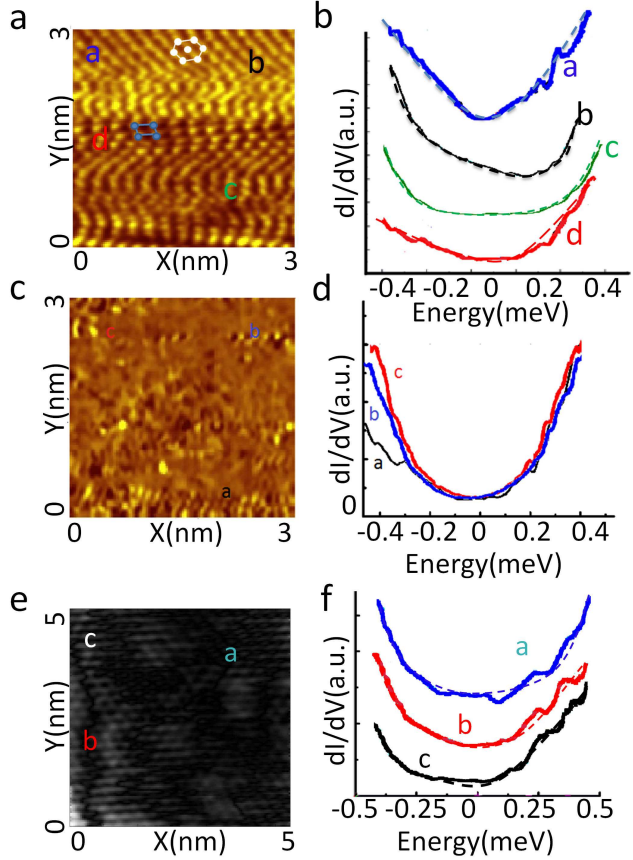}
\caption{(Color online) Lattice distortion of the CVD-grown graphene-on-copper samples and the resulting effect on the local electronic density of states: (a) A representative atomically resolved topography of one of the samples over a $(3.0 \times 3.0)$ nm$^2$ area, showing different atomic arrangements of the graphene sample in view, with slightly skewed honeycomb (or hexagonal) lattices in the $\alpha$-region and a combination of nearly squared and distorted atomic arrangements in the $\beta$-region. (b) Representative tunneling conductance $dI/dV$-vs.-$V$ point spectra in the $\alpha$-region (point 1) and $\beta$-region (points 1 and 2) shown in (a). The spectra are slighted shifted relative to each other along the $dI/dV$ axis for clarify. (c) A representative topography of the same sample in (a) over a disordered $(3.0 \times 3.0)$ nm$^2$ area. (d) Representative $dI/dV$-vs.-$V$ point spectra for the specific points in the $\gamma$-region shown in (c). (e) An atomically resolved topography of another CVD-grown graphene-on-copper sample over a $(5.0 \times 5.0)$ nm$^2$ area, showing distorted lattices. (f) Representative $dI/dV$-vs.-$V$ point spectra of the sample shown in (e). The spectra for points 1 and 2 are shifted up slightly along the $dI/dV$ axis for clarity. All data were taken at $T = 77$ K.}
\end{figure}

To better understand the significance of the conductance peaks, we subtract off a parabolic conductance background associated with the copper contribution (indicated by the dashed lines in Figs.~3b and 3f) and plot the resulting conductance against $(E-E_{\rm Dirac})$ where $E_{\rm Dirac}$ is the Dirac energy. As exemplified in Figs.~4a and 4c for the point spectra after background subtraction, we find distinct peaks occurring at energies proportional to $\sqrt{|n|}$ for $n = 0, \pm 1, \pm 2, \pm 3, 4$ and 6. Furthermore, for stronger strained $\beta$-type spectra, sharp peaks at energies proportional to fractional numbers $\pm \sqrt{1/3}$, $-\sqrt{2/3}$ and $\sqrt{5/3}$ are clearly visible, in contrast to the weak ``humps'' occurring in the weaker strained $\alpha$-type spectra at energies proportional to $\sqrt{1/3}$ and $-\sqrt{5/3}$. 

By plotting $(E-E_{\rm Dirac})$ versus $\sqrt{|n|}$ data for multiple spectra taken in Figs.~4a and 4c, we find the $(E-E_{\rm Dirac})$ data taken for each point spectrum in the atomically resolved strained regions all fall approximately on one linear curve, as exemplified in Figs.~4b and 4d, and the slope for each $(E-E_{\rm Dirac})$-vs-$|n|^{1/2}$ curve increases with increasing strain.    

\begin{figure}
  \centering
  \includegraphics[width=3.3in]{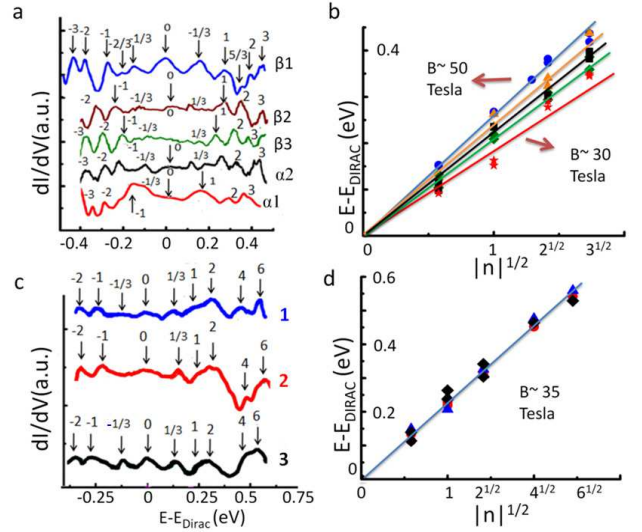}
\caption{(Color online) Manifestation of strained-induced pseudo-magnetic fields and quantized Landau levels in two CVD-grown graphene-on-copper samples at $T = 77$ K: (a) The tunneling conductance $(dI/dV)$ spectra vs. $(E-E_{\rm Dirac})$ after the subtraction of the parabolic conductance background shown in Fig.~3b. Here $E_{\rm Dirac}$ denotes the Dirac energy. Distinct peaks at energies proportional to $\pm \sqrt{|n|}$ are clearly visible for $n = 0, \pm 1, \pm 2, \pm 3$. Additional humps at fractional values $n = -5/3$, $-2/3$ and $\pm 1/3$ are also identifiable. (b) $(E-E_{\rm  Dirac})$-vs-$\sqrt{|n|}$ for the representative point spectra in (a), showing that the data largely fall on linear curves that correspond to pseudo-magnetic fields ranging from $B_s = 29 \pm 8$ Tesla for the less strained $\alpha$-region to $50 \pm 5$ Tesla for the stronger strained $\beta$-region. (c) The $(dI/dV)$ spectra vs. $(E-E_{\rm Dirac})$ after the subtraction of the parabolic conductance background shown in Fig.~3f. (d) $(E-E_{\rm  Dirac})$-vs-$\sqrt{|n|}$ for the representative point spectra in (c), showing a pseudo-magnetic field $B_s = 35 \pm 5$ Tesla.}
\end{figure}

The results manifested in Fig.~4 can be understood in terms of the presence of strain-induced pseudo-magnetic fields $B_s$, with larger strain giving rise to larger $B_s$ values~\cite{Guinea10a,Guinea10b,LevyN10}. Specifically, the pseudo-Landau levels $E_n$ of Dirac electrons under a given $B_s$ satisfy the following relation: 
\begin{equation}
E_n = {\rm sgn}(n) \sqrt{(2e v_{F} ^2 \hbar B_s)|n|},
\label{eq:En}
\end{equation}
where $v_F$ denotes the Fermi velocity of graphene, $B_s = |\nabla \times {\bf A}|$, and the strain-induced gauge field ${\bf A} = A_x \hat{x} + A_y \hat{y}$ is given by~\cite{Guinea10b}
\begin{eqnarray}
A_x &= \pm \frac{\beta}{a} (u_{xx} - u_{yy}), \qquad \qquad \qquad \nonumber\\
A_y &= \mp 2 \frac{\beta}{a} u_{xy}. \qquad \qquad \qquad \qquad \quad
\label{eq:Gauge}
\end{eqnarray}
In Eq.~(\ref{eq:Gauge}) $\beta \equiv - (\partial \ln t / \partial \ln a) \sim 2$, where $t \approx 3$ eV denotes the nearest-neighbor hopping constant, and $a$ is the nearest carbon-carbon separation. Additionally, the tensor components $u_{ij}$ ($i,j$ = $x,y$) are defined as $u_{ij} \equiv (\partial u_j / \partial i)$, where $u_j$ is the $j$th component of the displacement field ${\bf u} = u_x \hat{x} + u_y \hat{y}$.

Using $v_F \sim 10^6$ m/sec, we find that the pseudo-magnetic fields are approximately $29 \pm 8$ Tesla for the $\alpha$ region and $50 \pm 5$ Tesla for the $\beta$ region. These pseudo-magnetic fields are significant even though they are smaller than those found in graphene nano-bubbles~\cite{LevyN10}, probably due to the spatial inhomogeneity of the pseudo-magnetic fields that leads to partial cancellations. Further, the magnetic lengths $\ell _B$ associated with the pseudo-magnetic fields are estimated to range from 3.5 to 5.5 nm according to the following formula:
\begin{equation}
\ell _B ^{-2} = (2 \pi B_s/ \Phi _0), 
\label{eq:ellB}
\end{equation}
where $\Phi _0$ is the flux quantum. 

\begin{figure}
  \centering
  \includegraphics[width=3.3in]{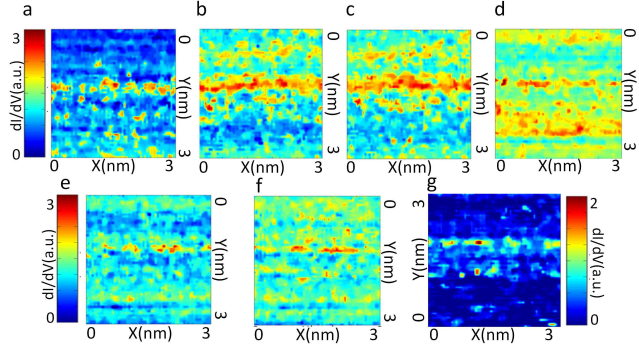}
\caption{(Color online) Constant-voltage tunneling conductance maps at quantized energies of the pseudo-Landau levels over the same ($3 \times 3$) nm$^2$ area and for $T = 77$ K: (a) $n = 0$, (b) $n = 1$, (c) $n = 2$, (d) $n = 3$, (e) $n = 1/3$, (f) $n = 2/3$, where the $n$ values are referenced to the strain-induced Landau levels for the $\beta 1$ point spectrum. For smaller $n$ values, an approximately one-dimensional high-conductance ``line'' appears near the topographical ridge where the most abrupt changes in height occur, suggesting significant charging effects. The confinement of high-conductance region diminishes with increasing $n$. (g) Spatial map of the Dirac-point tunneling conductance for $(dI/dV)(x,y)$ taken at spatially varying Dirac energy $V = (E_{\rm Dirac}/e)$.}
\end{figure}

The strain found in the CVD-grown graphene on copper is not purely shear but also contains compression/dilation components. The latter is theoretically predicted to gives rise to an effective scalar potential $V(x,y)$ and therefore a static charging effect~\cite{Guinea10b,SuzuuraH02,Manes07} in addition to the aforementioned pseudo-magnetic field. Here $V(x,y)$ is given by~\cite{Guinea10b,SuzuuraH02,Manes07}: 
\begin{equation}
V (x,y) = V_0 (\partial _x u_x + \partial _y u_y) \equiv V_0 \bar{u},
\label{eq:IVanti}
\end{equation}
where $V_0 \approx 3$ eV, and $\bar{u}$ is the dilation/compression strain. For strain induced by corrugations along the out of plane direction, the typical strain is of order $\bar{u} \sim (\Delta z/L)^2$, where $\Delta z$ denotes the height fluctuations and $L$ is the size of the strained region. Although the charging effect associated with the scalar potential can be suppressed in a graphene layer suspended over a metal if $\Delta z \ll \ell _B$~\cite{Guinea10a,Guinea10b}, the mean value of $\Delta z$ over the $(3 \times 3)$ nm$^2$ area of the CVD-grown graphene on copper was $\Delta z \sim 1$ nm, comparable to $\ell _B$. This quantitative consideration therefore suggests that significant charging effects may be observed.  

\begin{figure}
  \centering
  \includegraphics[width=3.3in]{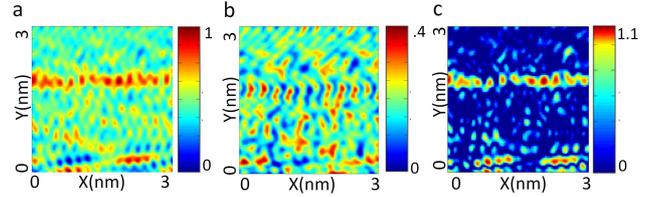}
\caption{(Color online) Spatial maps of the effective compression/dilation strain components for a ($3 \times 3$) nm$^2$ area of the CVD-grown graphene-on-copper sample shown in Figure 5, where the strain components are estimated by considering the spatial variations in the height $z (x,y)$: (a) $(\partial z/\partial x)^2$, (b) $(\partial z/\partial y)^2$, and (c) $\lbrack (\partial z/\partial x)^2 + (\partial z/\partial y)^2 \rbrack$.}
\end{figure}

To visualize the charging effect, we show in Figs.~5a - 5f the conductance maps for one of the CVD-grown graphene-on-copper samples at different constant bias voltages over the same $(3 \times 3)$ nm$^2$ area. The bias voltages are chosen so that they correspond to the pseudo-Landau levels of the $\beta 1$ point spectrum for $n = 0$, 1, 2, 3, 1/3 and 2/3. A nearly one-dimensional high-conductance region following closely to the most strained region immediately below the ridge is clearly visible in maps associated with smaller $n$ values, confirming the notion of strain-induced charging effects. For higher energies (larger $n$ values), the high-conductance region becomes less confined, which is reasonable because of the higher confinement energies ($\propto \sqrt{n}$) required for the Dirac electrons. We further note that the strain-induced charging effect also results in shifts in the Dirac energy. As manifested by the spatial map in Fig.~5g for the tunneling conductance taken at the spatially varying Dirac energy $E_{\rm Dirac} (x,y)$ for the same graphene sample, the $(dI/dV)(x,y)$ map for $V = (E_{\rm Dirac}/e)$ also exhibits excess conductance in strongly strained regions.  

To further manifest the compression/dilation strain-induced charging effect, we show in Figs.~6a - 6c spatial maps of effective compression/dilation strain components $(\partial z/\partial x)^2$, $(\partial z/\partial y)^2$ and $\lbrack (\partial z/\partial x)^2 + (\partial z/\partial y)^2 \rbrack$. Comparing Fig.~6c with the conductance maps in Fig.~5, particularly Fig.~5g, we find apparent correlation between the magnitude of strain and the conductance at the Dirac energy. 

\begin{figure}
  \centering
  \includegraphics[width=3.3in]{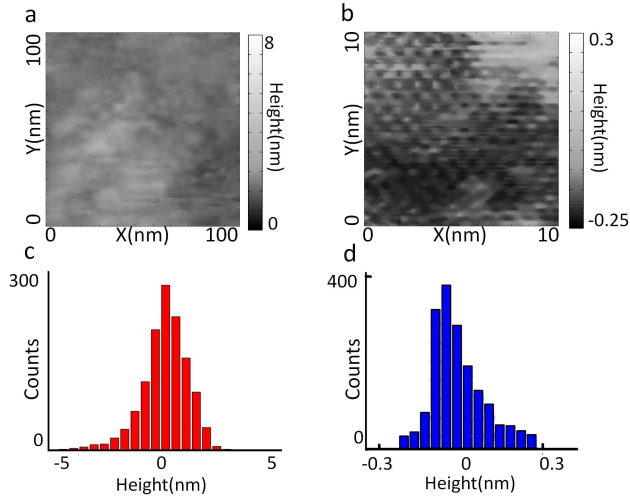}
\caption{(Color online) Topographic images of a CVD-grown graphene transferred from copper to SiO$_2$ for (a) a $(100 \times 100)$ nm$^2$ area and (b) a $(10 \times 10)$ nm$^2$ area. Both images were taken at 77 K. The corresponding height histograms for the images shown in (a) and (b) are given in (c) and (d), respectively. These height variations are much smaller than those found in Fig.~1a - 1c.}
\end{figure}

Next, we consider comparative studies of a CVD-grown graphene sample transferred from copper to SiO$_2$. As shown in Figs.~7a - 7d, the height variations in the transferred graphene sample are much reduced relative to those shown in Figs.~1a - 1c. The significant reduction in the average height variations of the ripples results in lattice structures and tunneling conductance spectra of the graphene sample similar to those found in mechanically exfoliated graphene on SiO$_2$~\cite{TeagueML09}, as exemplified in Figs.~8a and 8b. Moreover, the tunneling conductance spectra appear to be much smoother with no discernible conductance peaks in most of the transferred graphene-on-SiO$_2$ sample, as exemplified by the spectra in Figs.~8b and 8e for the points marked in Fig.~8a. On the other hand, occasional tunneling spectra near strained regions of the transferred graphene still revealed conductance peaks at quantized energies, as exemplified by the point spectrum in Figs.~8c and 8f. These regions contain topological ridges that cannot be relaxed by transferring the graphene sample from copper to SiO$_2$ substrate, so that the corresponding pseudo-magnetic field remains substantial, as exemplified in Fig.~8d where $B_s = 16 \pm 2$ Tesla.  

\begin{figure*}
  \centering
  \includegraphics[width=5.2in]{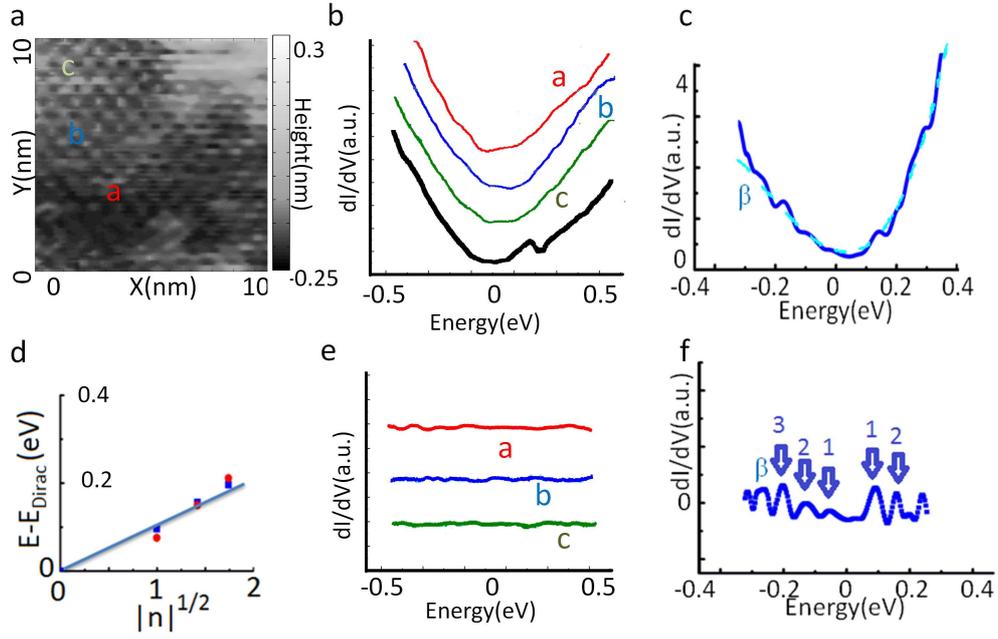}
\caption{(Color online) Topography and spectroscopy taken at 77 K on CVD-grown graphene transferred to SiO$_2$: (a) A representative topography over a $(10 \times 10)$ nm$^2$ area, showing gradual height variations and therefore corresponding to a relaxed region. (b) Representative tunneling conductance spectra of the relaxed region at multiple points marked in (a). The zero-bias conductance for the spectra taken at points a and b has been offset for easier comparison. Clearly all spectra are smooth without conductance peaks associated with pseudo-magnetic fields, in contrast to the findings for CVD-grown graphene-on-copper samples shown in Figs. 3 and 4. (c) A tunneling conductance spectrum in the transferred graphene-on-SiO$_2$ sample that corresponds a strained region (topography not shown). The spectrum reveals distinct quantized conductance peaks spaced at much smaller energies than those observed in Figure 4 for the CVD-grown graphene-on-copper samples. (d) $(E-E_{\rm Dirac})$-vs.-$\sqrt{|n|}$ for the spectrum taken in the strained region, showing that the conductance peaks fall on one linear curve that corresponds to $B_s = 16 \pm 2$ Tesla. (e) Multiple tunneling conductance spectra of the relaxed region subtracted by a background, showing no discernible conductance peaks and therefore implying $B_s \to 0$. (f) Tunneling conductance spectrum of the strained region subtracted by a background (indicated by the dashed line in (c)), showing conductance peaks at quantized energies, with $|n| = 1/3$, 1, 2 and 3.}
\end{figure*}

Similarly, the reduced strain for the transferred graphene-on-SiO$_2$ sample leads to reduced charging effect, as manifested by the overall lower and more spatially homogeneous Dirac-point conductance of the transferred graphene sample in Fig.~9c relative to that of the CVD-grown graphene-on-copper sample in Fig.~9a. Further comparison of the effective lattice distortion $\lbrack (\partial z/\partial x)^2 +(\partial z/\partial y)^2 \rbrack$ between the transferred graphene and the graphene-on-copper in Figs.~9b and 9d reveal that the compression/dilation strain is much reduced in the transferred sample. For strongly strained graphene, the magnitude of the compression/dilation strain correlates well with the Dirac-point conductance, as manifested by Figs.~9a and 9b. On the other hand, for negligible compression/dilation strain as in the case of transferred graphene-on-SiO$_2$, there is no longer direct correlation between the Dirac-point conductance and the local strain, because the former may be largely dependent on the local charge impurities on substrates. Overall, our STM/STS studies of various graphene samples have manifested the strong influence of strain and substrates on the structural rippling effect and the electronic DOS of graphene. 

\section{Discussion}
\label{sec4}

We have shown that significant differences between the thermal contraction coefficients of graphene and the transition-metal substrates for CVD growth graphene result in strong and non-uniform lattice distortion at the nano-scale, which provides a new ground for investigating the strain-induced charging and pseudo-magnetic field effects. In particular, the strain-induced giant pseudo-magnetic fields enable studies of the integer and fractional quantum Hall effects associated with the pseudo-spin ($i.e.$, the valley) degrees of freedom by exploiting the unique strength of the pseudo-spin-orbit coupling in graphene. This realization is analogous to the situation of two-dimensional topological insulators where a gapped bulk state coexists with the counter-propagating states at the boundaries without breaking the time-reversal symmetry~\cite{Kane05,Bernevig06,Moore07}. While in topological insulators the spin-orbit interaction is generally sizable ($10^{-1} \sim 1$ eV), in graphene the spin-orbit coupling is very small so that the corresponding Landau gaps are smaller than 1 $\mu$eV. Therefore, direct manifestation of the quantum spin Hall effect in graphene can only be realized under extreme empirical conditions of ultra-low temperatures ($< 10$ mK) and ultra-high carrier mobilities ($> 10^7$ cm$^2$ V s$^{-1}$). In contrast, the pseudo-magnetic field-induced quantum Hall effect associated with the pseudo-spins is easily identifiable in strained graphene, as demonstrated by this work and by similar studies of graphene nano-bubbles on Pt(111)~\cite{LevyN10}. 

\begin{figure}
  \centering
  \includegraphics[width=3.3in]{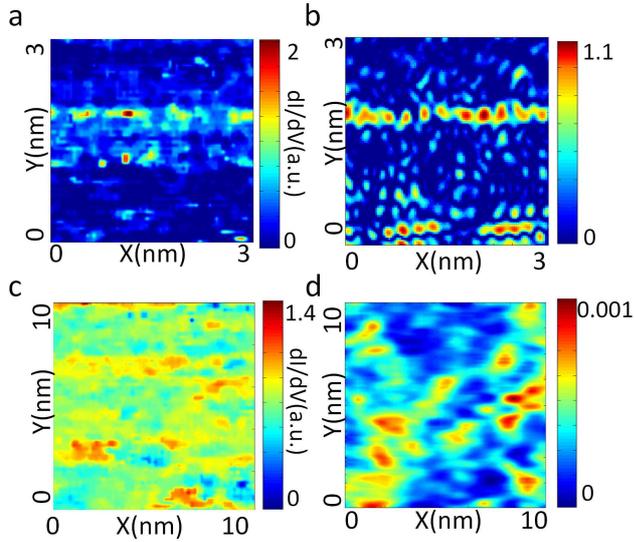}
\caption{(Color online) Strain effects on the Dirac-point tunneling conductance at $T = 77$ K: (a) A ($3 \times 3$) nm$^2$ spatial map of the Dirac point ($E = E_{\rm Dirac}$) tunneling conductance of the same CVD-grown graphene-on-copper sample as in Figure 3a. (b) Spatial map of the  compression/dilation strain $\lbrack (\partial z/\partial x)^2 +(\partial z/\partial y)^2 \rbrack$ over the same sample area as in (a). The maps in (a) and (b) are clearly well correlated. (c) A spatial map of the tunneling conductance of the transferred graphene-on-SiO$_2$ sample taken at the Dirac point $E_{\rm Dirac}$ over a ($3 \times 3$) nm$^2$ area. (d) The spatial map of the effective compression/dilation strain $\lbrack (\partial z/\partial x)^2 +(\partial z/\partial y)^2 \rbrack$ over the same sample area as in (c). The effective strain in (d) is several orders of magnitude smaller than that in (b). Hence, the lack of correlation between the maps in (c) and (d) implies that the charging effect in relaxed graphene is not directly due to strain and may be attributed to charge impurities on the substrate.}
\end{figure}

It is worth pointing out that our observation of pseudo-magnetic field-induced fractionally filled Landau levels in graphene may be attributed to the complicated strain distribution at the nano-scale, which ensures both large pseudo-magnetic fields and ballistic behavior of the Dirac electrons. Hence, strain-induced charge confinement in two dimensions (similar to the external magnetic field-induced charge confinement) gives rise to correlated many-body interactions. Moreover, the significantly more complicated strain distribution found in the CVD-grown graphene-on-copper differs from the controlled strain in the case of graphene nano-bubbles~\cite{LevyN10} where the effective gauge potential leads to a relatively uniform magnetic field over each nano-bubble. Therefore, further corrections to the Hamiltonian of the Dirac electrons are necessary in the case of CVD-grown graphene-on-copper, which may lead to additional terms in the effective gauge potential, similar to the presence of the Chern-Simons gauge potential terms in the case of fractional quantum Hall (FQH) phenomena. However, self-consistent theoretical studies will be necessary to account for the observation of FQH-like effects in the CVD-grown graphene-on-copper. 

We further remark that the tunneling spectra acquired from highly strained CVD-grown graphene-on-copper samples contain not only spectral characteristics of the strained graphene but also excess conduction channels associated with the underlying copper substrate. The excess conduction channels may be responsible for the commonly observed suppression of the zero-bias DOS associated with the $n = 0$ Landau level. However, the general effects of the underlying copper substrate on the graphene spectra are still unknown and await further theoretical investigation. 

The findings reported in this work are highly relevant to the development of graphene-based nano-electronics. In particular, proper ``strain engineering'' may be achieved by designing the substrate for graphene so that either local doping or energy-gap engineering associated with the pseudo-Landau levels can be realized at the nano-scale. Ultimately, strain-induced charging effects and pseudo-magnetic fields may also be achieved at the macroscopic scale if quasi-periodic occurrences of these nano-scale effects can be constructed.   

\section{Conclusions}
\label{sec5}
We have conducted spatially resolved topographic and spectroscopic studies of CVD-grown graphene-on-copper and transferred graphene-on-SiO$_2$ samples. Our investigation reveals the important influence of the substrate and strain on the carbon atomic arrangements and the electronic DOS of graphene. For CVD-grown graphene remaining on the copper substrate, the monolayer carbon structures exhibit ripples and are strongly strained, with different regions exhibiting varying lattice structures and electronic density of states. In particular, topographical ridges appear along the boundaries between different lattice structures, which also exhibit excess charging effects. Additionally, the large and non-uniform strain induces pseudo-magnetic fields up to $\sim 50$ Tesla, as manifested by the tunneling conductance peaks at quantized energies that are associated with both integer and fractional pseudo-magnetic field-induced Landau levels. In contrast, for graphene transferred from copper to SiO$_2$ after CVD growth, the typical graphene structure is largely restored and the average strain on the whole is much reduced, so are the corresponding charging effects and pseudo-magnetic fields, with the exception of limited areas that are associated with topological defects and therefore the resulting strain cannot be relaxed after the transfer of sample from copper to SiO$_2$ substrates. Our findings suggest feasible nano-scale strain engineering of the electronic states of CVD-grown graphene by proper design of the substrates and growth conditions.

\end{document}